# Glass clarified as the Self-Organizing System

**Elena A. Chechetkina**

*Institute of General and Inorganic Chemisctry of Russian Academy of Sciences (1980-2011)*

Moscow, Russia

eche2010@yandex.ru

*Abstract* – The term "self-organization" in a contemporary glass science relates usually to "intermediate phase" in frames of the *topological constraint theory* of glass structure. This theory, however, has no a relation to classical theory of self-organization – synergetics. The synergetic approach proposed here is based on characteristic instability of chemical bonding in the form of the *bond wave* considered as the spatiotemporal correlation between the elementary acts of bond exchange. In frames of the model, glass transition represents the *dimensionality* transition: from the 3D bond wave in glass-forming liquid to the 2D bond wave in glass. By using of available experimental data, the bond wave model displays *non-crystalline long-range order*, semi-deterministic behavior of viscosity, and *information field* as a necessary condition for glass formation.

## I. Introduction

Glass articles accompany us everywhere – from windows and kitchenware to fiber communications and active elements for electronic devices. At the same time no one material is so enigmatic, when even the nature of glass transition remains a subject of incessant discussions. To my opinion, the scientific mist appears because of ignoring of self-organization, a wide-spread phenomenon observed everywhere: in physics, chemistry, biology, psychology, sociology, etc. [1-3].

My understanding glass as the self-organizing system began in 1980s in the laboratory of *S. A. Dembovsky*, who was creating a new theory of glass formation based on the chemical bond point of view. The core of his theory are special bonds in the form of *hypervalent bonds*, which were considered initially as "defects" in covalent network [4] and after as a necessary element of glass structure, an element that ensures both glass properties and glass formation phenomenon at all [5]. During rummage of the "glassy" bonds by experiments of viscous flow in magnetic fields [6,7] my bond wave model was born.

*Bond wave* is considered as the spatiotemporal correlation between the bond transformation acts. The model was successively applied then for explanation of other glass features, both the known ones (medium-range order, non-Arrhenius behavior of viscosity, specific fracture) and those discovered by action of weak *information fields* (magnetic, ultrasound, etc.). The bond wave model history, present state and future development are considered n this paper.

## II. Dissipative Pattern – the Bond wave

Glass is undrstood usually as the *bulk non-crystalline material obtained by cooling from melt and then supercooled liquid up to complete solidification* [8,9]. Corresponding process is shown schematically in **Fig.1**, where $T_g$, $T_m$, $T_b$, and $T_m{}^*$ are the glass transition temperature, melting point, boiling point, and the "normal" melting point (see comments to **Table 1**).

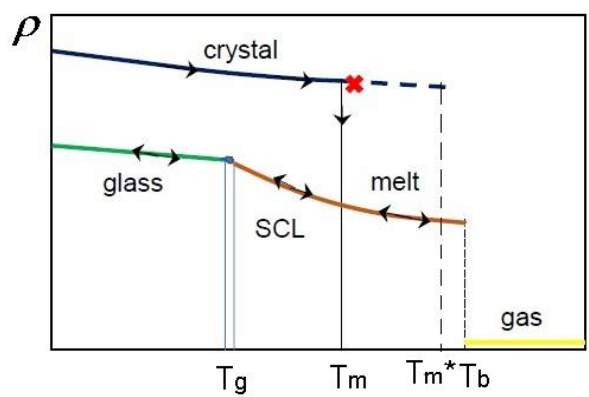

**Figure 1.** Density-temperature diagram for the crystal-melt-glass transformations; SCL is supercooled liquid.

Typical inorganic glasses belong to oxides ($SiO_2$, $B_2O_3$, etc.) or chalcogenides (Se, $As_2S_3$), both groups being based on **covalent bonds**, e.g., Si−O for $SiO_2$ or As−S for $As_2S_3$. Conventional **continuous random network** (**CRN**) model of glass structure [10] operates with the same covalent bonds in glass and crystal. For example, each O-atom of $SiO_2$ forms two covalent bonds with its neighbors, so being *two-coordinated* ($O_2$), and each Si-atom is *four-coordinated* ($Si_4$). This means that CRN operates with the **short-range order** (**SRO**), and there is no a difference between SRO in glassy and crystalline states of a substance.

Classical CRN (**Fig.2***a*) corresponds to the $B_2O_3$-like network consisting of $B_3$ and $O_2$ atoms. Such a CRN, however, cannot realize "in bulk" because of characteristic *rigidity* of covalent bond, which can change the bond length and valence angles only in narrow limits, a fact that makes classical CRN unstable after few random conjunctions.

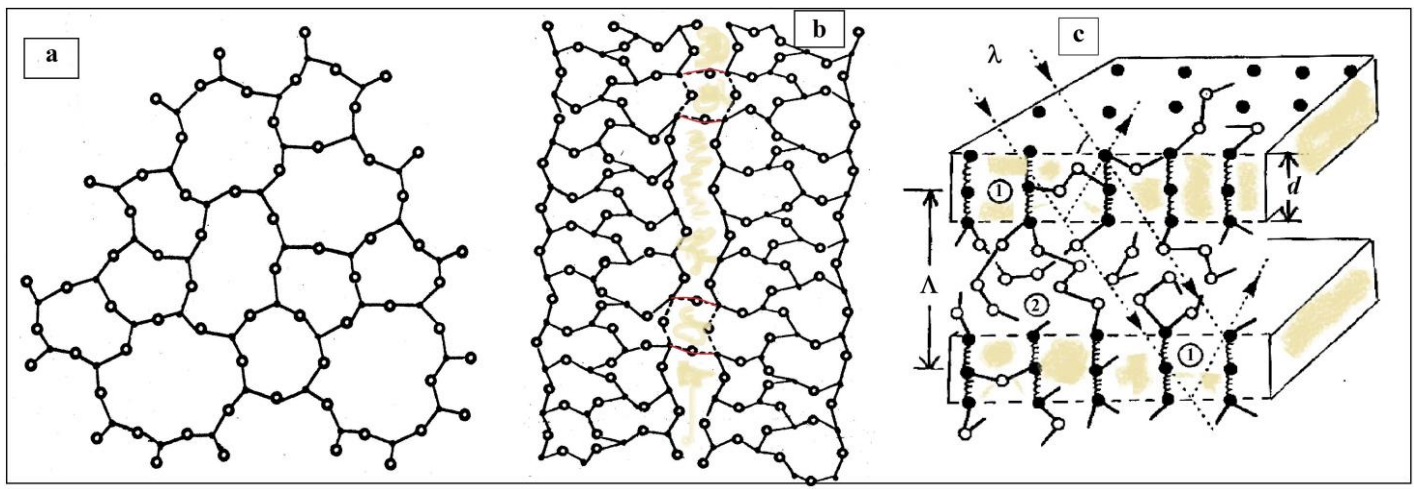

**Figure 2.** Models for glass structure after (a) *Zachariasen* [10], (b) *Robinson* [11] and (c) *Chechetkina* [12].

To overcome this shortcoming, *Robinson* [11] has introduced into CRN the "cuts" populated with weak/flexible bonds (**Fig.2b**), thus tailoring rigid covalent fragments with the following increase of CRN stability. Note that the notions about some "*weak bonds*" emerge periodically in glass literature (see, e.g., Ref.13); however, without appropriate justification of the bonds' nature.

Two other ways for reconstruction of classical CRN concern either special junction of covalent bonds such as "*outrigger rafts*" after *Phillips* [14] or the **topological constraint theory** of glass structure [15-17]. The theory considers only "normal" atomic coordination (i.e., $Si_4$, $O_2$, $As_3$, $Se_e$, etc.) which are used for calculation of the *average coordination number* $<r>$ in accord with chemical composition of a system. The obtained $<r>$ is compared with "*magic*" coordination number $<r>^*=2.4$ which corresponds to the free-of-stress state of a network. Really, the composition-property dependencies in binary glass-forming systems demonstrate anomaly in the region around corresponding "magic" composition; just this "*intermediate phase*" [18] is compared with the *self-organization* phenomenon in glass.

To my opinion, self-organization is a basic feature of the glassy state irrespectively of chemical composition. The bond wave model naturally combines "weak bonds" and "self-organization" owing to the *two-state* chemical bonding and a collective *feedback* between the states in the form of the **bond wave** representing the spatiotemporal correlation between elementary acts of reversible transformation between basic covalent bond (**CB**) and exited **alternative bond** (**AB**): $\Sigma\Sigma(CB \leftrightarrow AB)$, no matter now what is the AB nature.

The bond wave "elementary cell" is shown in **Fig.2c**, where alternative bonds are considered as the three-center bonds (TCB) in accord with the early model of hypervalent bonds after *Dembovsky* [4]. The following quantum-chemical study of "glassy" bond revealed concrete hypervalent bonds in typical glass formers (see review [5]); however, the nature of AB is of no matter now, as well as the network motive: 1D (like Se), 2D (like $B_2O_3$) or 3D (like $SiO_2$).

From a historical point of view, the bond wave model was born in the middle of 1980s owing to our magneto-viscous experiments [6,7], whose features – the field weakness and diamagnetic nature of glass – cannot be explained in frames of classical physics. Fortunately, the self-organization notions developed intensively at that time, but unfortunately, not in glass science, a fact that forced me to choose a roundabout way to meet glass community with the bond wave idea.

In doing so, I have used common thermodynamic data, phase transitions temperatures ($T_m$ and $T_b$) and atomization energy ($E_a$), for construction of the the $T_m=f(E_a)$ and $T_b=f(E_a)$ plots for "normal", molecular, and glass-forming substances [19]. Molecular substances, as expected, fall far below the "normal" lines. The fall of glass-formers is much less but remains strong enough to be neglected. For example:

**Table 1**. Atomization energy and phase transition temperatures for Se and $SiO_2$: real ($T_m$ and $T_b$) and "normal" ($T^*_m$ and $T^*_b$)

|        | $E_a$, kcal/g-at | $T_m$, K | $T^*_m$, K | $T_b$, K | $T^*_b$, K |
|--------|------------------|----------|------------|----------|------------|
| Se     | 52               | 494      | 940        | 958      | 1550       |
| $SiO_2$ | 149             | 1883     | 2700       | 2503     | 4400       |

This fact points to a substantial *inhomogeneity* in chemical bonding for two falling groups. Molecular substances are composed of strong covalent bonds within the molecules and very weak van-der-Waals bonds acting between them, so even a light heating leads to destruction of a molecular crystal; firstly at $T_m$, above which the temporarily closing molecules are linked with van-der-Waals bonds that switch from some atoms to others, and then at $T_b$, when van-der-Waals bonds break totally releasing free molecules into the gaseous phase.

The covalently bonded crystal endures thermal stress longer, up to a relatively high "normal" temperature $T^*_m$, above which the system of covalent bonds destroys totally because of metallization. Ideal metallic bond represents the positively charged frame united by the negatively charged cloud of electrons. Corresponding transition of bonding is observed by means of electric conductivity as the *insulator-to-metal* or the *semiconductor-to-metal* types of melting.

In contrast to "normal" crystals, the glass-forming ones does not metallize at melting. Let us consider the crystals of chalcogen group – S, Se, Te as a simple example. Tellurium, being the non-glass-forming substance, is known to metallize continuously when heating above $T_m$ [20], whereas glass-forming selenium and sulfur demonstrate the *semiconductor-to-semiconductor* transition [21] thus demonstrating conservation of covalent network in the molten state. This feature, which is known as the "polymeric" structure of glass-forming melts, does not explain, however, why covalent network becomes mobile after melting.

This puzzle can be resolved by means of alternative bonds and their wavelike self-organization as follows. A crystal of the glass-forming substance avoids metallization because of the low-temperature melting at $T_m<T^*_m$ (**Fig.1**, **Table 1**) owing to transformation of a piece of covalent bonds into the higher-energy alternative bonds accumulating excess thermal energy. Concentration of alternative bonds (AB) increases and reaches a critically high level when isolated can "feel" each other by local elastic fields generated in covalent network around each "alien" bond. For more and more effective accumulation of incoming thermal energy, alternative bonds gather into 2D layers, and then the layers organize into 3D bond wave representing the collectively moving layers populated with alternative bonds. Each step of AB self-organization means corresponding distortion of initial crystalline network, which destroys completely by 3D bond waves, whose wavefronts – the layers populated with alternative bonds continuously comb all the network in the wave direction.

Such a disappearance of crystalline order at melting does not mean, however, disappearance of order at all. Moreover, a rich *hierarchical* order appears. Let us illustrates this by means of **Fig2c** which presents three different orders. The scale of *short-range order*, **SRO**, is defined by the length of covalent bond shown by short lines between white circles/atoms; this length is $l=2.3$Å for Se (Se-Se), 1.7Å for SiO$_2$ (Si-O), etc. The SRO length can be extracted also from structural experiment as a radius of the *first coordination sphere*, $R_1 \approx 2$Å, which is obtained by Fourier transformation of diffraction pattern.

The diffraction pattern itself gives information about the *medium-range order*, **MRO**, observed as the *First Sharp Diffraction Peak* (FSDP). The MRO length corresponds to the wavefront thickness $d=2\pi/Q_1 \approx 4-6$Å, where $Q_1$ is the FSDP position. Equidistant wavefronts, two of which are shown in **Fig.2c**, give a relatively strong and *sharp* reflex. One can find other details about FSDP/MRO in my articles [12, 22-25].

The most intriguing is the *non-crystalline long-range order*, **NC-LRO**, in the form of Λ-lattice whose "elementary cell" one can see in **Fig.2c**. Unfortunately, Λ-lattice cannot be detected by ordinary X-ray analysis (λ=1-2Å) because of a strong parasitic scattering in the low-$Q$ region; the synchrotron radiation of varying λ is an alternative. Fortunately, a simple indirect observation of Λ-lattices is possible just now by means of fractography, as it will be demonstrated in Section V, after Section III and Section IV which considers theoretical (III) and experimental (IV) aspects of the bond wave model.

### III. ORDER PARAMETER – THE WAVELENGTH

Let us estimate the bond wave parameters as a function of temperature, beginning from concentration of alternative bonds

$$N = N_0 \cdot \exp(-\Delta\varepsilon/kT) \quad (1),$$

where $\Delta\varepsilon$ is the energy difference between the excited (AB) and the ground (CB) bonding states, and $N_0$ is a pre-exponent.

Of course, this approximation does not account interaction between alternative bonds, which, however, can be introduced by means of a simple geometry. From **Fig.2c** it follows

$$N/N_S = d/\Lambda \quad (2),$$

where $N_S$ is the AB concentration in $d$-layer.

By combining of Eq.(1) and Eq.(2), one obtains the temperature dependence of the bond wave **wavelength**

$$\Lambda = \Lambda 3 = d \cdot (N_S/N_0) \cdot \exp(\Delta\varepsilon/kT) \quad (3).$$

Note that **Λ3** relates to the *three-dimensional* bond wave corresponding to region 'I' on the top part of **Fig.3**; other regions will be considered some later.

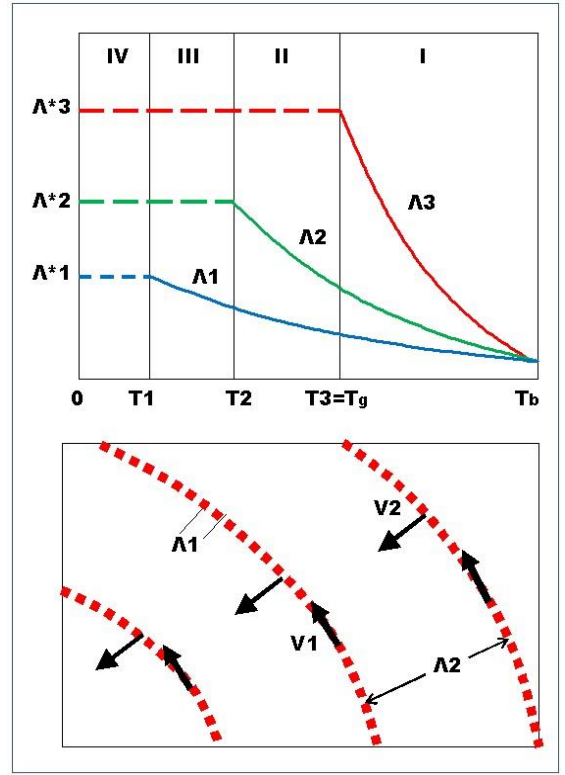

**Figure 3.** Temperature dependence of the wavelengths (top); 2D and 1D bond waves spreading along $d$-layer (bottom).

Bond wave dissipates thermal energy when moving through non-crystalline network with a temperature-dependent *velocity* **V**(*T*)=*f*(*T*)·*d*, where *f*(*T*)~exp(−$\varepsilon_f$/k*T*) is frequency of the temperature activated "jumps" (AB→CB→AB, etc.), *d* is the wavefront thickness (**Fig.2c**), and $\varepsilon_f$ is the jumping barrier. The reader can find numerical evaluation of **V**(*T*) for Se in Ref.26. Despite of the same *exponential* temperature dependence for the bond wave velocity and wavelength, **Λ**(*T*) and **V**(*T*), they change in the *opposite* way. In addition, the velocity changes with temperature faster than wavelength because $\varepsilon_f$>Δε (barrier is higher than related levels).

When melt *heating*, the wavelength of 3D bond wave decreases up to a critical temperature when the wavefronts become in contact (**Λ3**=*d*), so the network becomes homogeneous (*N*=$N_S$) and the bond wave disappears. This is a situation of **boiling**, when alternative bonds belonging to osculated wavefronts interact explosively with breaking and escaping of free covalent fragments into the gaseous phase.

In the opposite process of melt *cooling*, 3D bond wave reaches the second critical point, **T3**, when the critically distant wavefronts cannot "feel" each other by means of elastic field generated by them in surrounding covalent network. A feedback between the wavefronts breaks, and 3D bond wave stops, being *freezing* in the network in the form of **Λ*3**-lattice. As far as the *volume* mobility of a liquid is provided by the 3D bond wave, this freezing corresponds to **glass transition**; so **T3**=**T_g** and **Λ*3**=**Λ_g**, as it is shown at the top part of **Fig.3**.

The 2D and 1D bond waves, representing respectively collective/mobile "strings" in the limits of each *d*-layer and collective alternative bonds in the limits of each string, remains refrozen below **T_g**, as it is shown in the bottom part of **Fig.3**. Nevertheless, these low-dimensional waves also freeze at related temperatures **T2** and **T1**. Below **T1** glass is completely "dead" in the sense that every mobility provided by the bond waves is arrested there.

Fortunately, the low-temperature "death" of glass is *reversible* owing to thermal generation of alternative bonds and their ability for integration. When *heating*, there refreeze successively 1D bond wave at/above **T1**, the 2D bond wave at/above **T2**, and finally the 3D bond wave at/above **T3**=**T_g**. If the freezing/refreezing processes are not symmetrical (compare with braking/racing of a car), one can explain the phenomena of **hysteresis**, and not the well-known hysteresis around **T_g** (e.g., [8, p.29]) but also those at **T2** and **T1**. Corresponding two hysteresis cab be detected when studying the low-temperature properties and/or the so-called "secondary relaxation" in glass.

As a result, four temperature regions can be distinguished:

Region **I: Viscous liquid** (**T_g**-**T_b**). All three bond waves (1D, 2D, 3D) coexist although the region specificity is determined by 3D bond wave that animates all the volume and thus appears in "macroscopic" processes, *viscous flow* first.

Region **II: Plastic glass** (**T2**-**T_g**). Below **T3**=**T_g** concentration of alternative bonds in the frozen wavefronts remain high enough for realization of 2D bond waves which spreads along the stopped *d*-layers, thus providing 2D processes like *plastic flow* by sliding of *d*-layers under stress.

Region **III: Brittle glass**. Because of freezing of the 2D bond waves, glass enters the "brittle" region, when plastic flow and other 2D processes are arrested. A single response to mechanical stress is *destruction* of glass article, a fast process that denudes the frozen *d*-layers as the regions populated with a relatively weak alternative bonds; the jumping between the layers can be observed by characteristic *conchoidal fracture*.

Region **IV: "Dead" glass**. This hypothetical state is the most hard and immobile because of freezing of the bond waves of every dimension. Chaotic *explosive destruction* is expected.

## IV. ATTRACTOR FOR THE VISCOSITY-TEMPARATURE BEHAVIOR

Let us begin from **Region I** which corresponds to viscous liquid. The well-known feature of glass-forming liquids is the *non-Arrhenius* behavior of viscosity, which is seen in **Fig.4a** even for GeO$_2$, a typical "strong" liquid in terms of the strong-fragile classification after *Angell* [30].

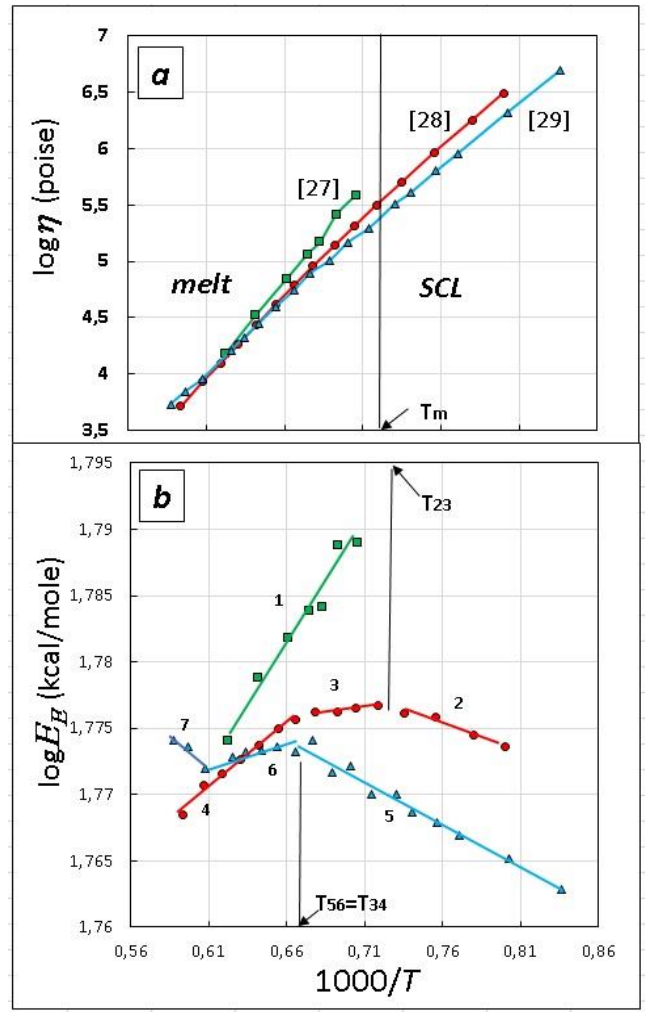

**Figure 4.** Experimental viscosity-temperature data for GeO$_2$ after *de Neufville et al* [27, Pt viscometer] (line 1), *Fontana & Plummer* [28, Run 1] (lines 2-4), and *Bruckner* [29] (lines 5-7) presented in the ***Arrhenius*** plot (***a***) and in the **TA-**plot (***b***).

Let call the *ideal strong liquid* a liquid that obeys the *Arrhenius* equation

$$\eta(T) = \eta_{Arr} \cdot \exp(E_{Arr}/RT) \quad (4)$$

with $\eta_{Arr}$=*const* and $E_{Arr}$=*const*. For a real liquid, the both parameters are temperature dependent. To reduce the number of variables, we have used earlier [31] the *Eyring* equation [32]

$$\eta(T) = \eta_E \cdot \exp(E_E/RT) \quad (5).$$

Here only $E_E$, the ***Eyring activation energy***, depends on temperature since the temperature dependence of $\eta_E = Nh/V$ ($N$ and $h$ are the Avogadro's and Plank's constants, and $V$ is the molar volume) can be neglected because the temperature dependence of density, $\rho \sim 1/V$, is negligible as compared with that for viscosity. Typical glass-formers considered below have practically the same pre-exponent $\log\eta_E \approx -4$.

In this way, common two-factor analysis of the viscosity-temperature data by Eq.(4) can be substitute for the one-factor $E_E = f(T)$ analysis irrespectively of the liquid fragility, as it was demonstrated by me earlier [33]. The analysis includes three steps of an increasing generality.

### 1. TA plot → Σ{$A_i$; $G_i$}

The first step begins from transformation of experimental $\eta(T)$ set into the $E_E(T)$ set by means of equations

$$E_E \text{ [kcal/mole]} = (4573/T) \cdot [\log\eta(T) - \log\eta_E] \quad (6)$$

and $\quad \eta_E$ [poise] $= 0.0039\rho$ [g/cm$^3$] / $M$ [g] $\quad (7)$,

where $\rho$ is average density and $M$ is molecular weight; $\log\eta_E$ is equal to −4.0 for GeO2, −3.8 for SiO2, −3.65 for Se, etc.

The $E_E(T)$ points then brings into the $\log E_E(T) = 1/T$ plot similar to that shown in **Fig.4b** for GeO2, which is actually the "*twice activation*" (**TA**) plot because it is *activation* plot for *activation* energy. The TA-plot not only emphasizes the distinction between experimental $\eta(T)$ data from different sources (e.g., [27-29] in **Fig.4a**) but also elicits a new natural law described by equation

$$E_{Ei} = A_i \cdot \exp(G_i/RT) \quad (8).$$

Thus obtained lines of the {$A_i$;$G_i$} parameters create a basis for the second analytical step.

### 2. Master plot → {$a$;$b$}

Although the arrangement of TA-lines in **Fig.4b** looks chaotical, the {$A_i$;$G_i$} pairs occur to relate as

$$\log A = a - b \cdot G \quad (9),$$

thus forming a *master line* like that shown in **Fig.5** for GeO2

The master line parameters, {$a$;$b$}, can be considered as **convergation point** whose coordinates defines all possible viscosity-temperature dependencies for the liquid considered. In terms of self-organization, convergation point represents the **attractor** for the viscosity-temperature behavior of a liquid.

In Table 2 there is a collection of convergation points for typical glass-formers, from strong to fragile, together with main characteristics of glass-forming liquids including the value of viscosity at melting point, $\log\eta(T_m)$.

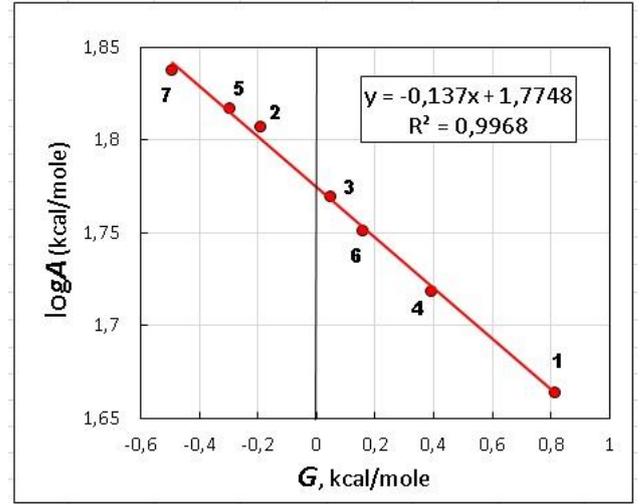

**Figure 5**. **Master plot** for GeO2. The point designation corresponds to the TA-lines in **Fig.4b**.

**Table 2.** Viscosity-temperature behavior for typical glass-forming liquids; fragility $m$ by Eq.(11), $\eta$ in [poise], $a$ and $b$ values are for [kcal/mole] dimension of $A$ and $G$ in Eq.(9).

| Substance | $T_g$, K $T_m$, K | $m$ $\log\eta(T_m)$ | $a$ $b$ | Refs for {$a$;$b$} |
|---|---|---|---|---|
| **SiO2** | 1500 1983 | 18 7.8 | **2.014** **0.101** | 28, 29 |
| **GeO2** | 880 1389 | 18 5.5 | **1.775** **0.137** | 27-29 |
| **B2O3** | 560 748 | 40 4.8 | **1.486** **0.298** | 34 |
| **Se** | 310 494 | 55 1.6 | **1.324** **0.700** | 35, 36 |
| **Gly** | 190 291 | 60 1.1 | **0.873** **0.779** | 37, 38 |

Based on the existence of convergation point, one can formulate the ***principle of partial reproducibility*** for the viscosity-temperature experimental data:
*viscosity of a liquid does not determined unambiguously by its temperature, the only one demand for the data correctness being their concordance with other measurements by fitting to the convergation point* {$a$;$b$} *for the liquid considered.*

This principle devaluates the ambition to obtain a "true" $\eta(T)$ dependence, as well as the attempts to construct the *general* equation for viscosity, a "Holy Grail" for generations of glass scientists. I propose instead the *universal* equation

$$\log\eta(T)_i = \log\eta_E + (A_i/2.303\,RT)\cdot\exp\{[(a-\log A_i)/b]/RT\} \quad (10)$$

with a fixed {$a$;$b$} and $A_i$ varying along the liquid master curve in discrete manner.

### 3. Convergation plot → Σ{*a*;*b*}

In **Table 2** it is seen that coordinates of convergation points change in the opposite direction and there is a relation between coordinates and the index of fragility defined as

$$m=|d(\log\eta(T)/d(T_g/T)|_{T_g} \quad (11).$$

These tendencies presented in **Fig.6**, distinguish two regions, the low-*a* region for "fragile" liquids (Se, glycerol) and the high-*a* region for "strong" liquids (GeO$_2$, SiO$_2$), "intermediate" B$_2$O$_3$ being gravitating to the "strong" group.

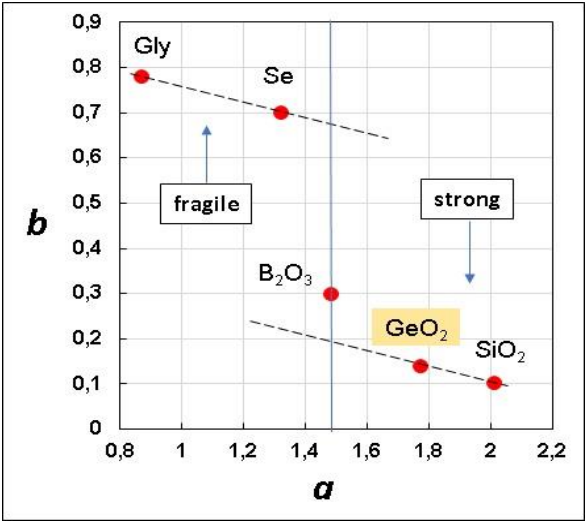

**Figure 6. Convergation plot** for the liquids in **Table 2**.

Investigation of convergation plot in more details will continue elsewhere, but now, in the context of self-organization, a possible mechanism by which a liquid selects one or another $\eta(T)_i$ from the permitted set is of interest. This is the problem of how adapts to its surrounding.

### V. INFORMATION FIELD FOR THE BOND WAVE

A general instrument for adaptation of a self-organizing system is *information*, by which a feedback between the system and its medium realizes [39, 40]. Bond wave needs information about *direction* to move, therefore, an **information field** that gives this direction is a necessary condition for glass formation. Really, if the bond wave does not "know" where to run it cannot realize, so formation of a low-temperature melt ($T_m<T^*_m$ in **Fig.1**) is impossible.

Viscous flow proceeds in the temperature region of refrozen 3D bond wave which, being a dissipative pattern that animates all the volume of a sample, is the most adaptive state of the glass-forming substance. A possible mechanism for adaptation of a sample adapts to flow conditions during a real experiment is shown in **Fig.7**, where a sample successively selects viscous pattern by means of the *feedback loops* between the "substance" and "information" columns. So even the cooling/heating mode or the run number are significant, a fact that is clear for best experimentalists (see, e.g., [37] or [28]).

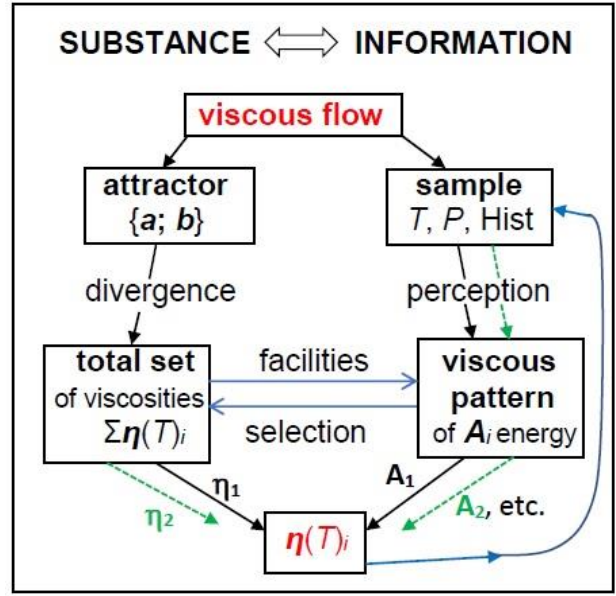

**Figure 7.** Self-organization at viscous flow of a sample depending on the sample chemical composition (substance), temperature (*T*), pressure (*P*), and history (*Hist*).

Although the main information field acting at viscous flow is the pressure gradient generated by viscometer (***grad*P**), other information fields can coexist during the process. For example, a combined action of two information fields, pressure and temperature gradients (***grad*P+*grad*T**), is shown in **Fig.8** using **fractography** as a simple method for observation of the frozen bond waves.

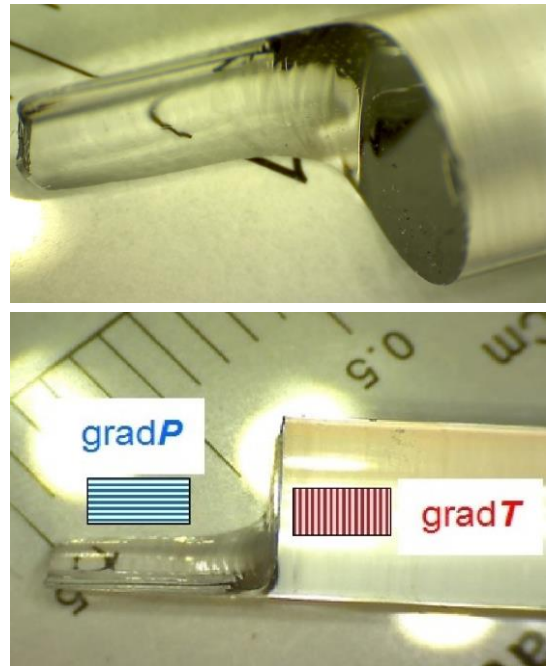

**Figure 8.** The fractured glassy rod (see text for details).

To draw a glassy rod from a softening ingot, one uses two gradients: the temperature gradient (*grad*T) from a heater and the pressure gradient (*grad*P) from a drawing device. The fracture in **Fig.8** elicits both waves by their Λ-lattices frozen in solid glass. A remarkable feature of this fracture is the *solitonic* behavior of underlaying bond waves, as far as solitons are defined as the waves intersecting without distortion. Although information fields were acting independently here, such behavior is not the case for our experiments listed in **Table 3**.

**Table 3**. The field effects studied in our laboratory; =**H** and ~**H** correspond to constant and pulsed magnetic field, **US** correspond to ultrasonic field in the cavitation regime.

| IF | Substance | Property | General effect | Ref. |
|---|---|---|---|---|
| =**H** 240Oe | Se | Viscosity (in situ) | $\Delta \lg \eta^H = \pm 0.2$ *Anisotropy* | [6] |
| ~**H** 240Oe, 50 Hz | Se | Viscosity (in situ) | *Resonance* $\Delta \lg \eta^H = +0.3/-0.5$ *Anisotropy* | [7] |
| =**H** 240Oe | $As_2S_3$ | Color  Fracture | *Red shift* for the "magnetic" sample (MS); *Plane* fracture (MS in **Fig.9**) | [41] |
| US 0.3 W/cm$^2$ | Se-Te | Optical transmission spectra.  SEM | *Non-linear* change in Se-X series of glasses; *Anisotropy*. Unusual images; Anisotropy (**Fig.11**; Se-Te) | [42] |
|  | Se-As |  |  | [43] |
|  | Se-S |  |  | [44] |
|  | Se:Cl |  |  | [45] |

First series of experiments on *viscous flow* of softening Se glass in **magnetic field [6,7]** was initiated by *Dembovsky* in his search for "glassy" bonds that provides switching of covalent bonds by means of charged intermediate states [4]. It was proposed that magnetic field can interact with electric current due to collective switching at viscous flow, so influencing on viscosity.

Although the expected effect in **constant** magnetic field was really observed [6], there were two problems that hampered the effect recognition. First, common glass like Se is the *diamagnetic* material – even the defects are considered to be diamagnetic in the basic state. Second problem is *weakness* of the applied field, whose energy is negligible as compered with thermal energy ($\mu_B H \ll kT$), so the magnetic field cannot interact with the proposed current, even if it exists.

Of course, the obtained effect would be presented as an experimental fact that needs further theoretical explanation, if not a relatively low value of the effect, $|\Delta \log \eta| \approx 0.2$, which can be easily attributed to experimental error. A simple way of increasing the field intensity for increasing the effect value was forbidden for us technically. Therefore, we changed the not the field intensity but the field character by using a **pulsed** magnetic field of the **50 Hz** frequency.

The pilot experiment at few temperatures showed no the magnetic field influence. However, after a detailed scanning of temperature, a new phenomenon of the **magneto-viscous resonance** was revealed, when in a narrow region at **321K** a strong effect, $|\Delta \lg \eta^H| \to 0.5$, was obtained [7].

To understand the magneto-viscous effects of both constant and pulsed fields, we have proposed first the existence of *one-electron* transition state B* which arises at the elementary act of bond exchange (CB↔B*↔AB) [46]. Then magnetic field interacts with not the previously proposed electric current but with paramagnetic states B* accompanying the process of switching of covalent bonds at viscous flow.

The *resonance* character of the effect indicates that frequency of thermally-activated acts of bond exchange coincides with the field frequency [46]. However, the problem of the field weakness remained. To overcome it, I have proposed that the B*-jumps are correlated not only in time by the temperature dependent frequency but also in space. In this moment the bond wave model was born, together with the information field concept.

In the magneto-viscous experiments there are two information fields, *grad*P and **H**. To consider the magnetic field only, the "magnetic//non-magnetic" pairs of Se, $As_2Se_3$ and $As_2S_3$ glasses were prepared in 1984 [47]. The case of $As_2S_3$ was especially interesting because of unusual color of the "magnetic" sample, which changed natural red color to almost black, as it is shown in **Fig.9**. The second difference has emerged ten years later, when I have splitting these samples for presentation in the MRS conference [41]. The cylindrical samples were splatted in the middle by a knife blade – the blow direction is shown by white arrows.

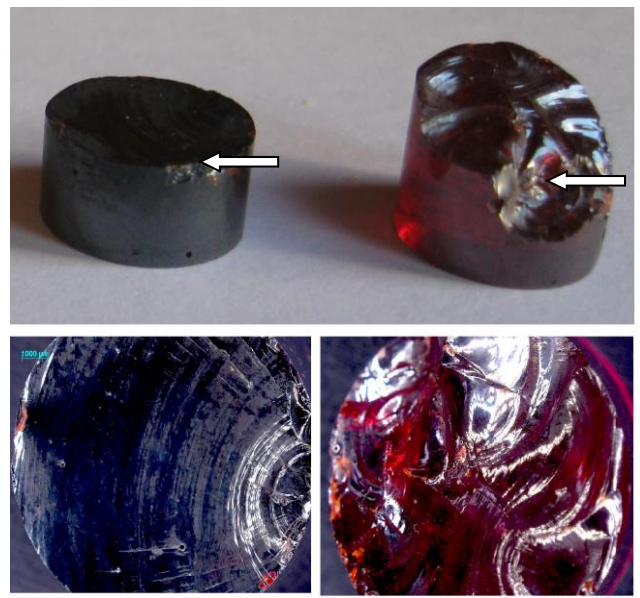

**Figure 9.** The fractured samples of "magnetic" (on the left) and "non-magnetic" (on the right) $As_2S_3$ glasses; a sight from sideway (top) and from above (bottom).

To understand both differences, one should consider the details of experiment. "Non-magnetic" and "magnetic" samples were prepared under the same conditions, namely, by melting of industrial $As_2S_3$ in evacuated quartz ampules at 450ºC for 1 hour in a tube furnace placed inside electromagnet, with the following 20 min cooling inside the cutting off furnace up to room temperature. The field of 240 Oe was acting on the "magnetic" sample *during the melting/cooling process.*

On the first glance, the features observed in "magnetic" sample conflict with the bond wave model. Really, as far as the *transversal* field of electromagnet orients the wavefronts *across* the tube/sample axis, the fracture should develop *along* the sample, in contrast to that observed in left-top corner of **Fig.9**. Besides, the darkening of "magnetic" sample remains a puzzle. The bond wave model resolves it when considering the bond wave dimensionality.

In accord with the top part of **Fig.3**, the bond waves of all three dimensions (1D, 2D, 3D) realize when treating the samples at temperatures above $T_g$. General dynamic factor here is the 3D bond wave which animates all the structure by its wavefronts representing *d*-layers populated with alternative bonds. As far as the acts of bond exchange through paramagnetic intermediate state are oriented by magnetic field, 3D bond wave runs in the field direction, and *d*-layers orient *along* the ampule/sample axis.

A critical situation appears when approaching $T_g$, when the 2D bond waves (the families of strings moving collectively along each *d*-layer – see the bottom part of **Fig.3**) become main dynamic factor. Now magnetic field acts mainly on the strings, which therefore ***turn*** their *d*-layers in the freezing network. The fracture observed in "magnetic" sample indicates that the strings have completed **rotation** of *d*-layers before total solidification of the sample.

The observed darkening of "magnetic" sample follows from this rotation. Those working in the field of glassy semiconductors are well-acquainted with the *red-shift* of optical edge observed after low-temperature illumination of chalcogenide amorphous films. We have proposed earlier an explanation of this phenomenon as a loosening of covalent network due to the light-induced drift of three-center bonds (hypothetical alternative bonds) [48]. By analogy, the red-shift observed in "magnetic" $As_2S_3$ can be explained by loosening of the solidifying covalent network due to the field-induced rotation of *d*-layers.

Finally, let consider a complex fracture of "non-magnetic" sample shown in the right part of **Fig.9**. The form of fracture reflects unstable thermal field $\Sigma grad_i T$ that was realizing during solidification. Then a plane cross-fracture of "magnetic" sample in **Fig.9** indicates that magnetic field has won the "information battle" between $\Sigma grad_i T$ and **H**.

Last experiment concerns **ultrasonic** (US) field, which was applied in the *cavitation* regime to softening Se-X glasses (X = Te, S, As, Cl) with Se as the main component. Fresh samples displayed a strong ***non-linearity*** observed in the form of extremum on the concentration-property curves at 1-2%Te, 5%As, 1%S, and 0.01-0.02%Cl [49]. Owing to a low glass-transitions temperature of the samples ($T_g \approx 40$ºC for Se), a simple cell (**Fig.10**) with water as cavitation medium was used. The treatment temperatures are 40ºC, 50ºC, and 72ºC; the treatment time is some minutes (2-5 min depending on temperature); the field intensity is about 0.3 W/cm$^2$.

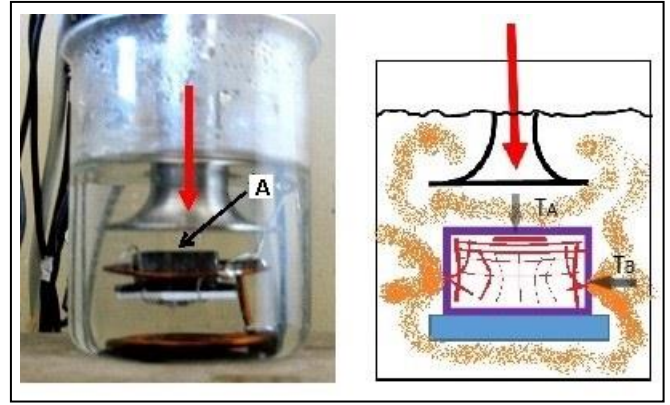

**Figure 10**. US-treatment of Se-X samples: the cell (on the left) and the scheme of interaction (on the right). Red arrow is the US input, red curves are the wavefront section. Arrows $T_A$ and $T_B$ indicate two directions for measurement of optical spectra.

The samples of a given Se-X series were measured by optical transmission spectra in the 300-5000 cm$^{-1}$ range before and after each treatment. Specific lines and *transparency*, $T$, (optical transmission at the 1000 cm$^{-1}$, a frequency that divides vibrational region and the "window of transparency" for selenide glasses) was watched during experiment. Since there was no change in spectra of Se standard hold in water of 72ºC for 5 minutes when the US transmitter was turned off, one can attributed the observed effects just to the US influence.

Let us consider the most remarkable features of the cavitation effects including the non-linear response using **Fig.11**. There are two final samples from the same Se-Te series; the fractures are parallel (**B**) or perpendicular (**A**) to the US-input indicated in **Fig.10** by thick red arrows.

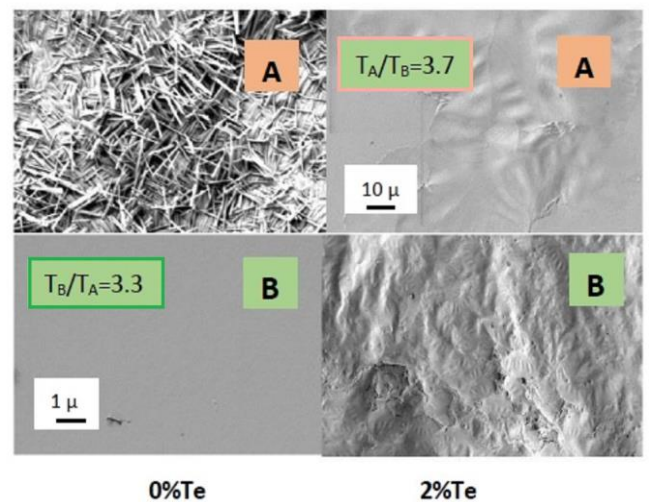

**Figure 11.** SEM images of fractured Se and $Se_{98}Te_2$ samples.

Two samples in **Fig.11** belong to the 0-1-2-5-10%Te series, the 0%Te (pure Se) being a "normal" member while the 2%Te being an "extremal" one because it fits in the 1-2%Te region of non-linearity. Transparency became very low after final 72°C treatment, nevertheless, the measured values indicate clearly *optical anisotropy* and, moreover, the anisotropy *inversion*: from $T_A<T_B$ for "normal" Se to $T_A>T_B$ for "extremal" 2%Te. Note that before US-treatments all glasses were normally isotropic ($T_A=T_B$) and of the same transparency.

Although all fractures looked glassy by the eye (smooth and bright), the SEM images were not, except lateral fracture of pure Se (**B**-fracture for 0%Te in **Fig.11**). The difference in transparency of the samples measured in frontal ($T_A$) and lateral ($T_B$) directions ($T_B/T_A$=3.3 for 0%Te and $T_A/T_B$=3.7 for 2%Te) coincides with their SEM images, when the observed inhomogeneities act as light scatters. The problem is their unusual form. The *needles* on the **A**-fracture of Se (0%Te) look like microcrystals, however, Se in known to crystallize in the *spherulitic* manner. The *clouds*, which are observed on the both fractures of 2%Te sample, look neither crystalline nor glassy, being probably an intermediate *glass-to-crystal* form arising due to cavitation treatment of this "extremal" sample.

The problem is that we do not know *when* the observed "needles" and "clouds", were formed: *during* cavitation treatment or *after*, when the samples were cooling and/or when keeping before SEM. The last variant has emerged due to repeat SEM of Se (0%Te) after 6 months: the needles have appeared on the **B**-fracture too, although not as a carpet but as a family of needle islands. This shows that the cavitation-induced crystallization continues in solid glass, which therefore remains in an activated state even below $T_g$. In every case, the *low-temperature* ($T<T_g$) crystallization, as a process provided by 2D bond waves, should differ from the *high-temperature* ($T>T_g$) crystallization realized by means of 3D bond waves. The difference is not only in kinetics but also in morphology and in the volume-or-surface location [50].

## VI. DISCUSSION

The proposed approach to glassy state differs from conventional one in basic points indicated in **Table 4**. Let us consider the difference in frames of four topics usually discussed by glass community – from structure and properties to glass transition and practical usage.

**1. Structure**. Common definition of glass as "a solid having no long-range order" means not more than "glass is not a crystal" when common *crystal*-type long-range order (LRO) is meant as usual. Bond wave creates a specific **non-crystalline LRO** that coexists naturally with the medium-range order (MRO) and the short-range order (SRO) [51]. Corresponding scales are the wavelength $\Lambda$ for NC-LRO, the wavefront thickness $d$ for MRO, and the covalent bond length $l$ for SRO (see **Fig.2*c*** for visualization).

When cooling of supercooled liquid at/below $T_m$, there appears a *competition* between 3D bond wave supporting NC-LRO and a tendency to establish "normal" crystal-type LRO. Formation of glass means that 3D bond wave has won the battle of orders.

**Table 4.** Glass paradigms reconsidered.

| GLASS | Conventional notions | The Bond Wave model |
|---|---|---|
| **Long-Range-Order** | Absent | NC-LRO in the form of $\Lambda$-lattice |
| **Chemical Bonding** | One-state: covalent bond | Two-state: CB↔AB |
| **Reproducibility at $T>T_g$** | Total, depending on temperature | Partial, depending on glass history |
| **Glass Transition** | Various models and theories | 3D BW → 2D BW transition at $T_g$ |
| **Glass Structure** | Models/theories based on CRN | Hierarchical: SRO/ /MRO/NC-LRO |
| **Management of Properties by...** | …Chemical Composition | …Composition and Information Fields |
| **Glass Definition** | Non-crystalline solid obtained by cooling from melt at a sufficiently high rate (critical cooling rate) | Non-crystalline solid obtained from a liquid possessing self-organization ability in the form of the bond waves |

*Abbreavitures*: CB – covalent bond, CRN – continuous random network, NC-LRO – non-crystalline long-range order, AB – alternative bond, BW – bond wave, MRO – medium-range-order (known also as IRO – intermediate range order).

**2. Reproducibility**. Supercooled liquid that exists in the $T_g<T<T_m$ range is assumed to be in the "metastable equilibrium", a term that implies a definite property-temperature relation. Then every appreciable deviation from a "true" property-temperature dependence is scarified as "experimental error". It was demonstrated in Section IV that the viscosity-temperature behavior of glass-forming liquids does not determined strictly, being possessing a *freedom* which is restricted only by **convergation point** of the liquid – or **attractor** in terms of self-organization.

One may expect similar attractors for other temperature dependent properties. As to the viscosity attractor, it generates a family of *viscous patterns* which represents themselves as lines in the twice-activation coordinates. The observed transition between the patterns at $T_{ij}$ can be compared with the "liquid-liquid" phase transitions (see [52] for introduction), although this is a far analogy for at least two reasons. First, the transition between viscous patterns looks rather chaotic (like 1-7 lines in **Fig.4b**), being depending not only on temperature but also on the sample history (**Fig.7**). Second, if one interprets the $\eta(T)_i$ viscous pattern as a "phase", this is the *dynamic* phase transition observed under *normal pressure*. The pressure-induced shift of convergation point will be considered in my next paper.

**3. Glass Transition** is a central point in glass science, being also an object of a permanent discussion for more than century. Main disagreement concerns the nature of a dramatic loss of mobility below $T_g$, the glass transition temperature. In frames of the bond wave model, glass transition is the *dimensionality* one: from 3D BW in liquid to 2D BW in solid. As far as 3D bond waves are frozen/stopped below $T_g$, the *volume* mobility, including viscous flow, is arrested below $T_g$, where only the low-dimension processes, such as plastic flow along the stopped *d*-layers, remain. This is my answer to the principal question "*why glasses do not flow*" [53]. As to the numerous theories/models of glass transition proposed so far, I know the only one that also uses dimensionality – this is the *Ojovan* model for silica which consideres a hypothetical system of percolating clusters named "vitrons" [54]. This model, however, leads to the *increase* of dimensionality when cooling – from fractal dimensionality $d_f$=2.3 in liquid to Euclidian dimensionality $d$=3 in glass.

**4. Practice**. It is quite evident that development of technology needs a deeper understanding of the processes implicated. With the bond wave model, there appears a possibility for reasonable usage of the low-energy *information fields* to improve the glass making process. Note that information fields just present at the process, at least in the form of temperature and pressure gradients (**grad***T*+**grad***P*), and one can use them more effectively based on the underlaying bond wave picture.

Besides common temperature/pressure gradients, one can introduce other information fields, e.g., *magnetic* and *ultrasound* fields (**H** and **US**), whose influence on softening bulk glasses was demonstrated in Section V. Applicability of the bond wave model to thin films was demonstrated in Ref.55 on the example of memory elements based on the electric switching phenomenon (*electric* information field) and microlenses formed under laser illumination (*electromagnetic* information field).

When using information fields, one should take in mind two implications. First, external field can provide both *energy* – for the bond wave support, and *information* – for giving the wave direction. A simple example is the thermal field of a given intensity (temperature, the energy level) and direction (temperature gradient, information). Second, there are usually two or more potential information fields, which can *coexist* in the solitonic manner (see **grad***T* and **grad***P* in **Fig.8**), *interact* (as **grad***P* and **H** in our magneto-viscous experiments), and *depress* a competitor (**H** has eliminated **grad***T* in the "magnetic" sample – **Fig.9**). Besides, a post-action of information field is possible (see the last comment to **Fig.11**).

## VII. CONCLUSIONS

A new approach to glassy state, which combines classical *self-organization* and the *chemical bond* theory in the context of "glassy" bonds after *S. A. Dembovsky* (1932-2010), is proposed. This approach is realized in the form of the **bond wave model** which, with the use of available experimental data, leads to nontrivial notions about (1) *non-crystalline long-range order*, (2) *partial reproducibility* of experimental data, and (3) *information fields* as a necessary condition for glass formation and a perspective instrument for management of glass properties.

The bond wave model concentrates now on the wavefronts, i.e., on a relatively small (but exceptionally active!) piece of non-crystalline structure. As far as the wavefronts pass continuously through every point/atom of a network, the *"secondary" self-organization* develops in the most of structure; in this way a link between the "*topological*" and the "*bond wave*" aspects of self-organization appears.